\documentclass{ws-ijmpd}
\usepackage[super,compress]{cite}

\begin{document}


%
\catchline{}{}{}{}{}
%

\title{CENTRIFUGAL  ACCELERATION IN THE ISOTROPIC PHOTON FIELD
}

\author{G. G. BAKHTADZE
}

\address{School of Physics, Free University of Tbilisi\\
Tbilisi, 0159, Georgia\\
gbakh15@freeuni.edu.ge}

\author{V. I. BEREZHIANI}

\address{School of Physics, Free University of Tbilisi\\
Tbilisi, 0159, Georgia
\\
Andronikashvili Institute of Physics (TSU), Tbilisi 0177, Georgia
\\
v.berezhiani@freeuni.edu.ge}

\author{Z. OSMANOV}

\address{School of Physics, Free University of Tbilisi\\
Tbilisi, 0159, Georgia\\
z.osmanov@freeuni.edu.ge
}

\maketitle

\begin{history}
\received{Day Month Year}
\revised{Day Month Year}
\end{history}

\begin{abstract}
In this paper we study centrifugal acceleration of particles moving along a prescribed rotating curved trajectories. We consider the physical system embedded in an isotropic  photon field and study the influence of the photon drag force on the acceleration process. For this purpose we study three major configurations of the field lines: the straight line; the Archimede's spiral and the dipolar field line configuration. By analysing dynamics of particles sliding along the field lines in the equatorial plane we have found several interesting features of motion. In particular, it has been shown that for rectilinear field lines the particles reach the light cylinder (area where the linear velocity of rotation exactly equals the speed of light) zone relatively slowly for bigger drag forces. Considering the Archimede's spiral, we have found that in cases when the field lines lag behind the rotation, the particles achieve the force-free regime of dynamics regardless of the drag force. Unlike this scenario, when the spiral is oriented in an opposite direction the particles do not reach the force free regime, but tend to stable equilibrium locations. Such a behaviour has been found for straight field lines as well.
\end{abstract}


\ccode{PACS numbers: 98.70.Sa; 97.60.Lf; 98.62.Js}

\section{Introduction}
%
%
%
%

 According to observations the very high energy (VHE) emission in the TeV band (Ref. \refcite{aleksic}) and cosmic ray particles from TeV (Ref. \refcite{electrons}) to PeV (Ref. \refcite{pev}) and even much higher (Ref. \refcite{abraham}) are detected by several modern facilities  and therefore, one of the major problems in this context is to understand the origin of such energetic cosmic rays. Despite many attempts this problem still remains the long standing one modern astrophysics encounters.

 In order to explain enormous energies of cosmic rays a prominent physicist Enrico Fermi has proposed a process, which now is called Fermi acceleration mechanism. According to his approach the cosmic particles are accelerated in the interstellar space by means of collisions against moving magnetic walls (Ref. \refcite{fermi}). A substantial modification has been considered in (Refs. \refcite{bell1,bell2}), but as it has been realised this Fermi-type processes are efficient only for "pre-accelerated" particles with relatively high initial energies (Ref. \refcite{rieger}).

 An interesting alternative mechanism has been hypothesised  by Gold for explaining the pulsating radio sources (Refs. \refcite{gold1,gold2}). The author has argued that due to the frozen-in condition plasma particles follow rapidly rotating magnetic field lines and as a result, undergo very efficient centrifugal force, leading to efficient acceleration of particles, which after gaining enormous energy, of the order of TeV,  can radiate away VHE photons. A very simplified theoretical model has been studied in (Ref. \refcite{mr}) where the authors have considered dynamics of particles freely sliding along straight rotating channels. It has been shown that by the influence of relativistic effects of rotation the particles might efficiently accelerate in a region called light cylinder - a hypothetical zone, where linear velocity of rotation exactly equals the speed of light. Although in realistic astrophysical situations the field lines are often curved. Therefore, a certain generalisation of the mentioned model has been presented in (Ref. \refcite{rdo}), where instead of straight rotating channels the curved trajectories has been studied. It is worth noting that the detected cosmic rays (Refs. \refcite{electrons}-\refcite{abraham}) have already left a region where they originated from, therefore, it is clear to assume for them the force-free regime. By taking into account this fact, the autors have found that in order for particles to reach the mentioned dynamical pattern the channels must have the shape of the Archimede's spiral. On the other hand, one cna show that in the light cylinder zone, by means of the curvature drift instability the magnetosphere of pulsars and active galactic nuclei (AGN) reconstruct and the field lines twist behind the rotation getting the shape of the Archimede's spiral  (Ref. \refcite{reconstr1,reconstr2}). A series of articles has been dedicated to mechanism of magnetocentrifugal acceleration in the magnetospheres of pulsars and AGNs (Ref. \refcite{applic1}-\refcite{applic5}).

Generally speaking, when charged particles accelerate, they radiate, inevitably leading to the so-called radiation reaction force (see Ref. \refcite{berezh} and references therein), that in turn, will influence the overall dynamics of particles. This particular problem has been studied in (Ref. \refcite{drb}), where the authors have shown that the radiation reaction force under certain conditions may lead to centripetal motion. It also has been found that due to the existence of radiation reaction force locations of stable equilibrium might exist. Similar but a slightly different physical situation arises when centrifugally accelerating particles move in a photon background and as a result undergo the so-called radiation drag force. Such a physical situation is very natural because nearby zones of many extreme astrophysical objects are characterised by high photon field densities.Therefore, it is interesting to consider this particular problem and study how centrifugal acceleration is influenced by the photon drag force.

The paper is organised in the following way. In section~II, a theoretical model of centrifugal acceleration is presented, in section~III we consider results and in section ~IV we summarise them.

\section{Theoretical model} \label{sec:main}
We consider particle motion in two dimensions alongside a wire anchored to the central object which is rotating with angular velocity $\omega$ . The form of the wire is given in polar coordinates as $\varphi(r)$ and consequently the channel is located in the equatorial plane. In general, magnetospheres of astrophysical objects are full of a photon field which we consider to be isotropic in the laboratory reference frame and in certain cases its contribution in dynamics of particles might be significant. This particular term might be given as a friction force acting from the high density photons on the moving particles
\begin{equation}\label{drag}
  \mathbf{F_f} = - \beta \gamma^2 \mathbf{v},
\end{equation}
where $\gamma$ is the Lorenz factor of the particle, $\mathbf{v}$ is its total velocity in the laboratory reference frame (LF), $\beta$ is expressed as to be (see Ref. \refcite{berezh} and references therein)
\begin{equation}\label{beta}
 \beta = \frac{4}{3c}\sigma_{_T}U
\end{equation}
$c$ is the speed of light and $U$ is the energy density of the photon field.

After following a method developed in (Ref. \refcite{rdo}) and taking into account that the channels are co-rotating with angular velocity $\omega$, the effective angular velocity in the LF writes as
\begin{equation}\label{omef}
 \Omega(t)=\omega+\varphi' \dot{r},
\end{equation}
which leads to the following Lorenz factor of the particle
\begin{equation}\label{gama}
 \gamma(t)=\frac{1}{\sqrt{1-\dot{r}^2-\Omega^2 r^2}},
\end{equation}
where $\dot{r}\equiv dr/dt$ and henceforth we use $c=1$.

The radial and azimuthal components of particle's relativistic momentum are given by
\begin{eqnarray}
  P_r (t) &=& \gamma m \dot{r} \label{pr},\\
  P_\phi (t) &=& \gamma m \Omega r \label{pfi}.
\end{eqnarray}
After introducing the reaction force, $\mathbf{N}$, acting on the particle from the field line, one can straightforwardly check that the set of equations governing the dynamics of a particle writes as
\begin{eqnarray}
  \frac{d(\gamma m \dot{r})}{dt}-\gamma m \Omega^2 r &=& -N\sin\alpha - \beta \gamma^2 \dot{r} \label{dpr} \\
  \frac{d(\gamma m \Omega)}{dt} r + 2\gamma m \Omega \dot{r} &=& N\cos\alpha - \beta \gamma^2 \Omega r \label{dpfi}
\end{eqnarray}
where $\alpha = \arctan{(r \varphi')}$ is the angle between the radius vector of the particle and the tangent to the field line at particle's location. Eq.~(\ref{dpr}) and Eq.~(\ref{dpfi}) finally reduce to the following expression for the radial acceleration (for the detailed derivation see Ref. \refcite{rdo})
\begin{equation}\label{a}
 \ddot{r}=\frac{r \Omega \omega - \gamma^2 r \dot{r} (\omega \dot{r} + \varphi') (\Omega + r \dot{r} \varphi'') }{(1-\omega^2 r^2+r^2 \varphi'^2) \gamma^2} - \frac{\beta (\dot{r} + r^2 \varphi' \Omega)}{(1-\omega^2 r^2+r^2 \varphi'^2) \gamma m}.
\end{equation}
It is worth noting that for translating the dimensionless values into physical values, one has to restore true dimensions of velocities by $v\rightarrow v/c$, $\omega r\rightarrow\omega r/c$.

\section{Discussion} \label{sec:res}
%
%
%
%

In this section we study the influence of the photon field drag force on dynamics of particles. For this purpose we examine several particular configurations of the field lines: Straight line, Archimedes' spiral and the Dipolar field line.

\subsection{Straight line}

\begin{figure}
  \centering
  \includegraphics[width=\textwidth]{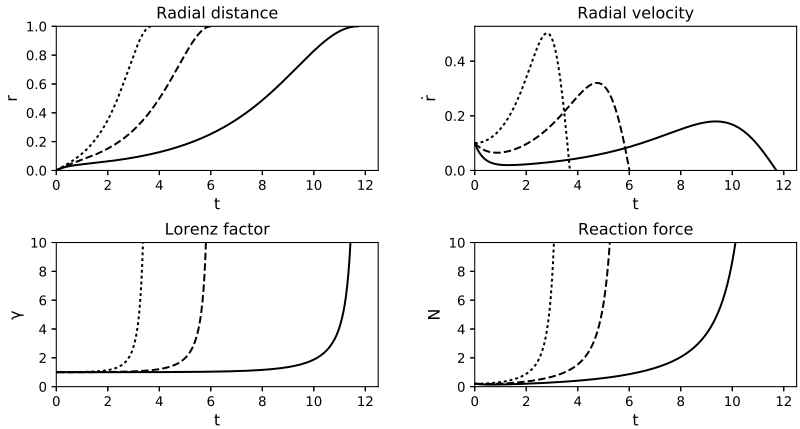}
  \caption{Plots for the radial distance $r(t)$, the radial velocity $\dot{r}(t)$, the Lorenz factor $\gamma(t)$ and the reaction force $N(t)$. The set of parameters is $m=1$, $\omega = 1 $, $r(0) = 0$ and $\dot{r}(0) = 0.1 $, $\beta = 0.0$ (dotted line), $\beta = 1.0$ (dashed line) and $\beta = 2.5$ (solid line). }\label{fig1}
\end{figure}

\begin{figure}
  \centering
  \includegraphics[width=0.9\textwidth]{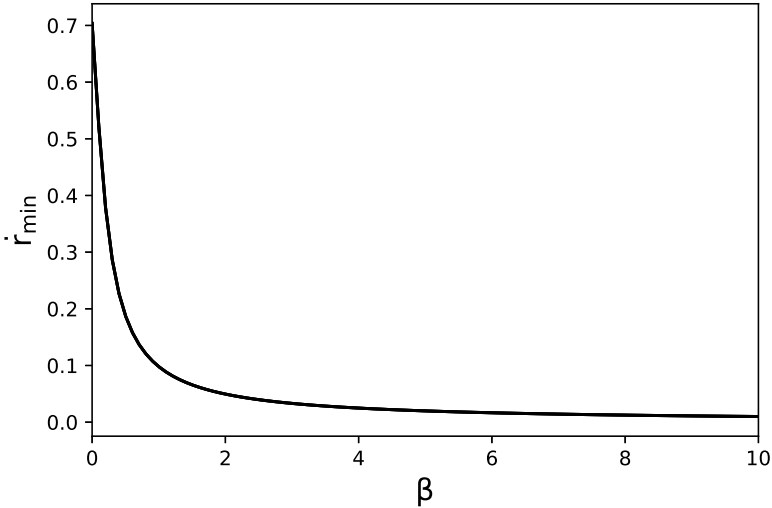}
  \caption{Dependence of minimum initial radial velocity versus $\beta$. Shape of wire is straight line. The set of parameters is: $\omega = 1$, $m = 1$, $r(0) = 0.1$}\label{fig2}
\end{figure}

In this case the shape of a wire is given by $\varphi = const$, which leads Eq.~(\ref{a}) to the following expression
\begin{equation} \label{a1}
  \ddot{r}=\omega^2 r - \frac{2 \omega^2 r \dot{r}^2 }{1-\omega^2 r^2} - \frac{\beta \dot{r}}{\gamma m (1-\omega^2 r^2)}
\end{equation}
For numerical solutions we need two initial conditions. Those are initial radial distance $r(0)=r_0$ and initial radial velocity $\dot{r}(0) = v_0 $. It is worth noting that motion of a particle for $\beta>0$ is a bit different from a scenario when there is no friction ($\beta = 0$). Results are shown in Fig.~\ref{fig1}, where we plot temporal evolution of the radial coordinate, radial velocity, and the reaction force. The set of parameters is $m=1$, $\omega = 1 $, $r(0) = 0$ and $\dot{r}(0) = 0.1 $, $\beta = 0.0$ (dotted line), $\beta = 1.0$ (dashed line) and $\beta = 2.5$ (solid line). As it is clear from the plots by increasing the photon drag force, motion slows down in time. As a result, the particle reaches the light cylinder, but for higher values of $\beta$ this happens later. We see the similar behaviour for the Lorenz factor and reaction force. As it is clear from these results, the corresponding timescale is higher for high density photon field, implying that efficiency of acceleration is less for the mentioned case. On the other hand, the accelerated particles will inevitably radiate in the presence of magnetic field and therefore the radiation pattern will be strongly dependent on the process of energy gain.
\newline
From the previous study (see Ref. \refcite{drb}) it is known that with no friction if at the origin initial velocity is more than $1/\sqrt{2}$ then the radial acceleration becomes negative from the very beginning of motion. The situation changes if friction is present. As as it is clear from the plots, the initial acceleration is negative if $\beta>0$. For the general case, when the photon field is present, from Eq.~(\ref{a1}) one can straightforwardly show that the minimum initial radial velocity, for which the initial radial acceleration is negative is given by
\begin{equation} \label{vmin}
  \dot{r}_{min} (\beta) = \left[\frac{1-\omega^2 r_0^2}{2} \left(1-\sqrt{1-\left[1+\left(\frac{\beta}{2 \omega^2 r_0 m}\right)^2\right]^{-1}}\right)\right]^{\frac{1}{2}}
\end{equation}
it is evident that when $\beta > 0$ and $r_0 \rightarrow 0$ then $\dot{r}_{min} \rightarrow 0$. In Fig.~(\ref{fig2}) we show the behaviour of $\dot{r}_{min} (\beta)$. We see that the minimum initial velocity is a continuously decreasing function of the drag coefficient. This particular dependence might have a certain impact on the acceleration mechanism in astrophysical situations. In particular, as it is clear from the results, radial motion becomes more restricted in high photon field density media and as we will see in the following subsections, this might lead to stable equilibrium in certain locations of a magnetosphere, which might be observationally evident.

\subsection{Archimede's spiral}
In this subsection we consider a particular configuration of field lines: the Archimede's spiral $\varphi = a r$ where $a$ is constant. Then Eq.~(\ref{a}) reduces to
\begin{equation}\label{a2}
  \ddot{r}=\frac{r \Omega \omega - \gamma^2 r \dot{r} (\omega \dot{r} + a) \Omega }{(1-\omega^2 r^2+r^2 a^2) \gamma^2} - \frac{\beta (\dot{r} + r^2 a \Omega)}{(1-\omega^2 r^2+r^2 a^2) \gamma m}.
\end{equation}
It is particularly interesting to examine two principally different cases of the Archimede's spiral: a) $a<0$ and $a>0$.

\subsubsection{Archimedes spiral with $a<0$}

\begin{figure}
  \centering
  \includegraphics[width=\textwidth]{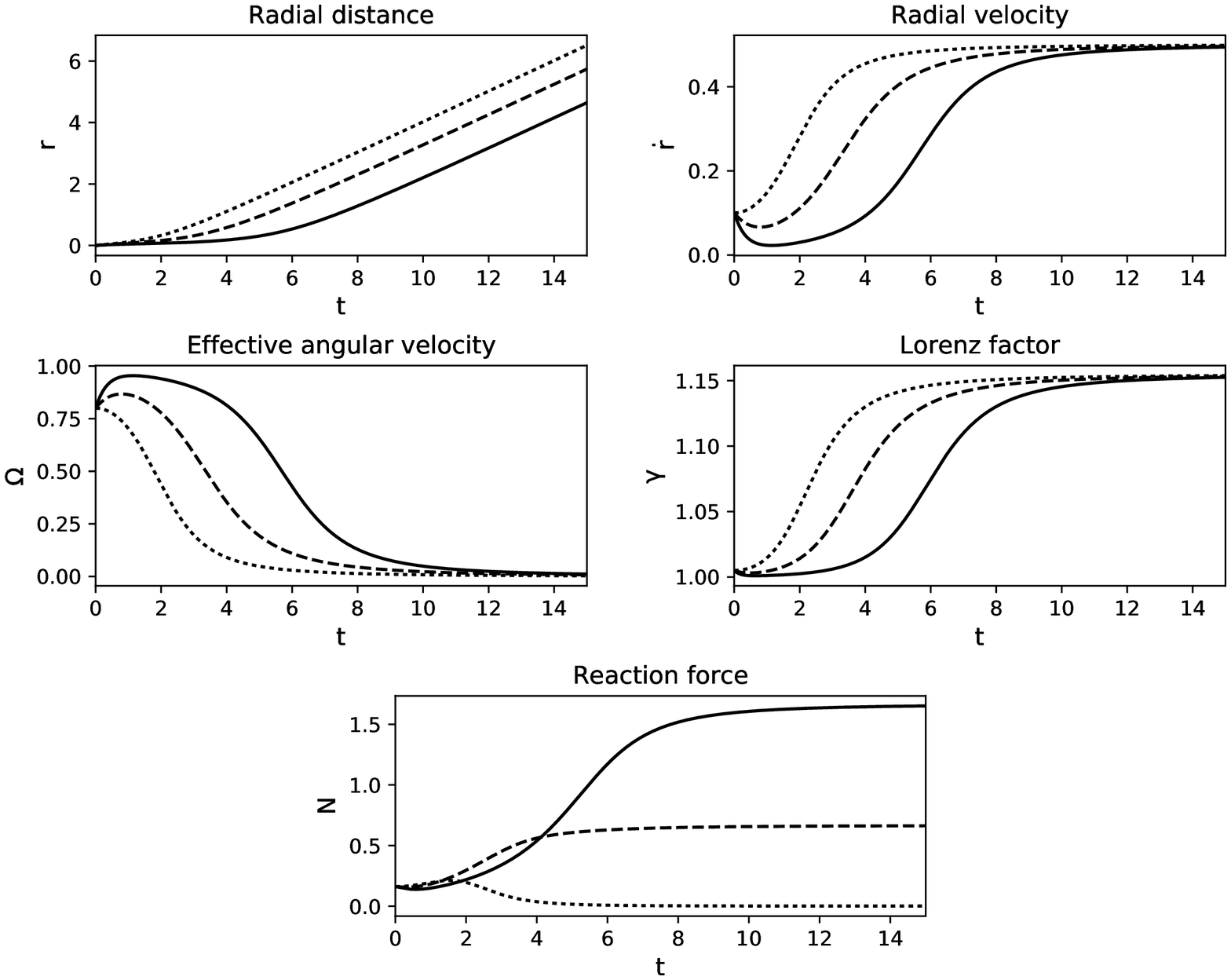}
  \caption{Plots for the radial distance $r(t)$, the radial velocity $\dot{r}(t)$, the effective angular velocity, $\Omega(t)$, the Lorenz factor $\gamma(t)$ and the reaction force $N(t)$. The set of parameters is $m=1$, $\omega = 1 $, $r(0) = 0$, $a=-2.0$, $\dot{r}(0) = 0.1$, $\beta = 0.0$ (dotted line), $\beta = 1.0$ (dashed line) and $\beta = 2.5$ (solid line).}\label{fig3}
\end{figure}

\begin{figure}
  \centering
  \includegraphics[width=\textwidth]{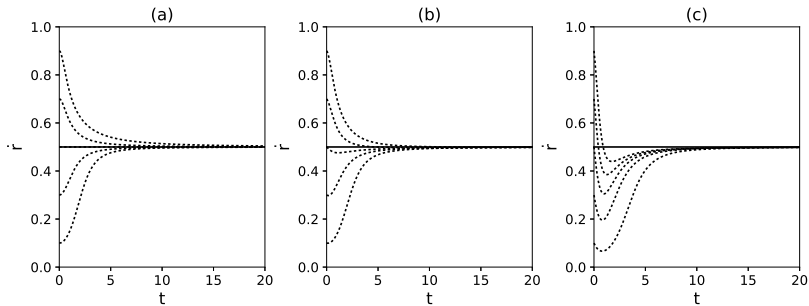}
  \caption{Plots of radial velocity versus time for different initial values.$r(0)=0$, $\omega = 1.0$, $a=-2.0$. (a) $\beta = 0.0$, (b) $\beta = 0.1$, (c) $\beta = 1.0$. Solid line shows asymptotic value of radial velocity ($- \omega/ a$)}\label{fig4}
\end{figure}

Under certain conditions dynamics of the particle tends to be force free. As it is seen from Fig.~(\ref{fig3}), regardless of the value of $\beta$ effective angular velocity decreases and asymptotically vanishes, but the process becomes relatively slow for bigger values of $\beta$. As a result, from Eq.~(\ref{omef}) we obtain $\dot{r} \rightarrow -\omega / a$.

In Fig.~(\ref{fig4}) we present the temporal behaviour of radial velocity for different initial velocities and different photon field drag coefficients. The set of parameters is $r(0)=0$, $\omega = 1.0$, $a=-2.0$, (a) $\beta = 0.0$, (b) $\beta = 0.1$, (c) $\beta = 1.0$. The horizontal solid line corresponds to the asymptotic radial velocity. As we see, the radial velocity tends to the corresponding asymptotic value, thus the dynamics becomes fore-free. The similar result is clearly seen in the plot of $\Omega(t)$ (see Fig.~(\ref{fig3})). It is worth noting that force-free dynamics for the negligible photon field on the one hand and the significant drag force on the other is achieved differently. In particular, for $\beta = 0$ the value of reaction force vanishes (Ref. \refcite{rdo}), but when $\beta > 0$ the corresponding reaction force tends to the asymptotic value of the friction force canceling it. One can straightforwardly show that for the reaction force one has
\begin{equation}\label{Nas}
  N \rightarrow \frac{\beta \omega a}{\omega^2 -a^2}.
\end{equation}
Therefore, the saturated value of $N$ is proportional to $\beta$. One more difference is that if in non-friction mode with the initial radial velocity $-\omega / a$ it always remains the same, for $\beta > 0$ and the same initial conditions the particle starts to decelerate at first and then tends to this value (see Fig.~(\ref{fig4})). It is clear that $-\omega / a$ must be less than $1$ in order to achieve the force free regime, otherwise dynamics is much like the case $a=0$.

\subsubsection{Archimedes spiral with $a>0$}

\begin{figure}
  \centering
  \includegraphics[width=\textwidth]{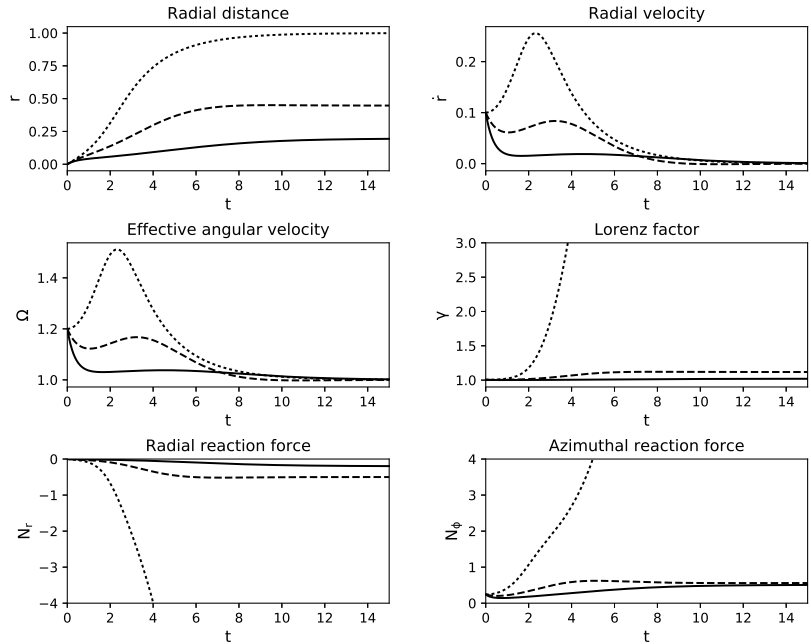}
  \caption{Plots for the radial distance $r(t)$, the radial velocity $\dot{r}(t)$, the effective angular velocity, $\Omega(t)$, the Lorenz factor $\gamma(t)$ and the components of the reaction force $N_r (t)$, $N_\phi (t)$. The set of parameters is $m=1$, $\omega = 1 $, $r(0) = 0$, $a=2.0$, $\dot{r}(0) = 0.1$, $\beta = 0.0$ (dotted line), $\beta = 1.0$ (dashed line) and $\beta = 2.5$ (solid line).}\label{fig5}
\end{figure}

\begin{figure}
  \centering
  \includegraphics[width=0.9\textwidth]{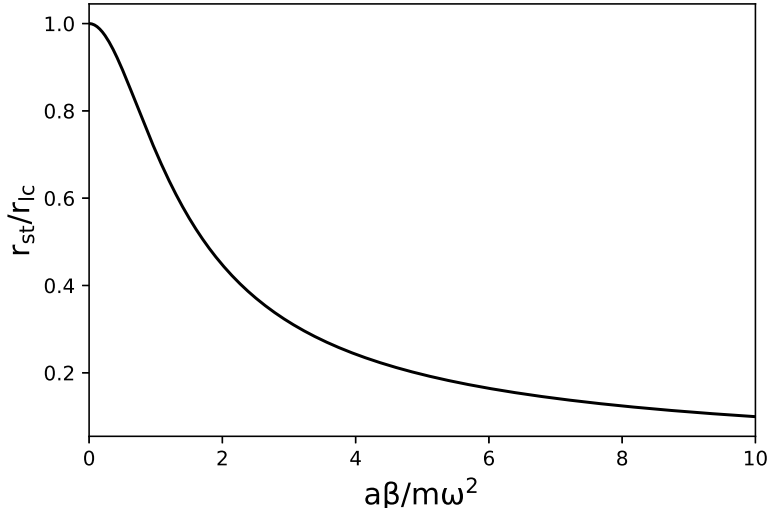}
  \caption{Dependance of the ratio $r_{st}/r_{lc}$ on the parameter $a \beta / m\omega^2$}\label{fig6}
\end{figure}

This case is completely different from the previous one. Configuration of the channel is such that the particle can be in stable equilibrium. Numerical results show that regardless of the value of $\beta$ particle's radial velocity and radial acceleration both tend to zero and consequently the radial distance tends to a finite value. This means that effective angular velocity finally becomes $\omega$. This is also clearly seen from numerical solutions (see Fig.~(\ref{fig5})). After substituting $\dot{r}=0$ and $\ddot{r}=0$ into Eq.~(\ref{a}) one can derive an expression for the stable radial distance
\begin{equation}\label{rst}
  \frac{m \omega}{\beta} \sqrt{1-\omega^2 r_{st}^2} = r_{st} \varphi'(r_{st}),
\end{equation}
being valid for any $\varphi(r)$. For the particular case $\varphi(r) = a r, a>0$ this equation has one solution
\begin{equation}\label{rst1}
  r_{st}=\frac{1}{\sqrt{\omega ^2 + \left( \frac{a \beta}{m \omega} \right)^2}}.
\end{equation}
It clear that $r_{st}$ is less than the light cylinder radius if $a$ is positive and the photon drag force is present. Notice that ratio of the stable radial distance in the light cylinder coordinates depends only on the following dimensionless parameter $a \beta / m\omega^2$
\begin{equation}\label{rst2}
  \frac{r_{st}}{r_{lc}}=\frac{1}{\sqrt{1 + \left( \frac{a \beta}{m \omega^2} \right)^2}},
\end{equation}
where $r_{lc} = 1/\omega$ is the light cylinder radius. The corresponding behaviour is shown in Fig.~(\ref{fig6}).
From Eqs. (\ref{dpr}) and (\ref{dpfi}) it is evident that the reaction force components are given by
\begin{eqnarray}
  N_r &=& -\frac{m}{\sqrt{1-\omega^2 r_{st}^2}} \omega^2 r_{st} \label{Nr}.\\
  N_\phi &=& \beta \frac{1}{1-\omega^2 r_{st}^2} \omega r_{st} \label{Nfi}.
\end{eqnarray}
These expressions  also follow from the fact that all the forces at stable equilibrium point must be zero. Radial component must be equal to centrifugal force $\gamma m \omega^2 r_{st}$, and azimuthal to friction force $\beta \gamma ^2 \omega r_{st} $.

\subsection{Magnetic field line}
\begin{figure}
  \centering
  \includegraphics[width=\textwidth]{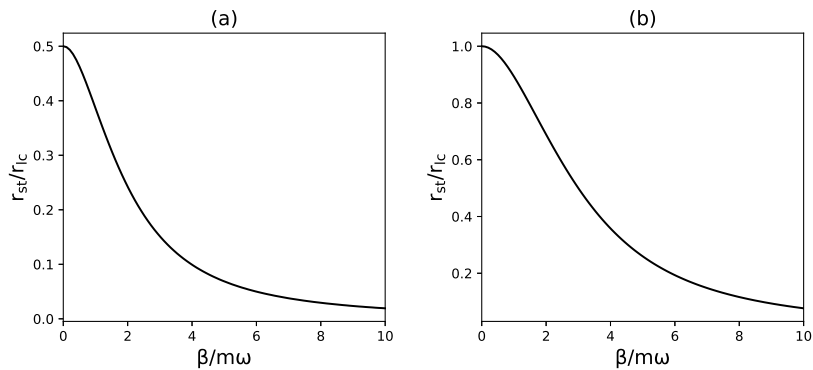}
  \caption{Dependence of the ratio $r_{st}/r_{lc}$ on the parameter $\beta / m \omega$ when for two different values of the parameter $L/r_{lc}$: (a) 0.5 and (b) 2.0}\label{fig7}
\end{figure}

In this subsection we study stable equilibrium points on a field lines corresponding to the dipolar magnetic field. The shape of the magnetic field lines is given by the equation $r = L \sin^2 \varphi$ where L is constant and defines a length scale of a field line. When the dipole magnetic moment is located in the equatorial plane, which is the case we consider here, then $\varphi(r)$ can be expressed as follows
\begin{eqnarray}
  \varphi_1(r) &=& \arcsin{\sqrt{\frac{r}{L}}} \label{fi1},\\
  \varphi_2(r) &=& \pi - \arcsin{\sqrt{\frac{r}{L}}} \label{fi2},
\end{eqnarray}
where $r$ changes from zero to $L$. It is obvious that stable equilibrium point must be on $\varphi_1 (r)$ because $\varphi'_2 (r)$ is negative and consequently Eq.~(\ref{rst}) does not have solution. Therefore equation for $r_{st}/r_{lc} = s$ is given by
\begin{equation}\label{rst3}
  \frac{2 m \omega}{\beta} \sqrt{1-s^2} = \sqrt{\frac{s}{\frac{L}{r_{lc}}-s}}.
\end{equation}
Its obvious that solution for $s$ depends only on two parameters: $\beta / m \omega$ and $L / r_{lc}$. Although this equation can be transformed to a cubic form, only one of three solutions will satisfy Eq. (\ref{rst3}). The corresponding numerical solutions are given in Fig.~(\ref{fig7}). As one can see from the plots as $\beta \rightarrow 0$, $r_{st}$ either approaches $r_{lc}$ or $L$. That depends or relation of $r_{lc}$ and $L$. If $L / r_{lc} < 1$ (the field line is fully inside the light cylinder)  when $\beta \rightarrow 0$, $r_{st}$ approaches to $L$ . If $L / r_{lc} > 1$ (the field line is only partially inside the light cylinder)  as $\beta \rightarrow 0$, $r_{st}$ approaches the light cylinder area. As $\beta$ increases $r_{st}$ vanishes in the both cases.

\section{Summary} \label{sec:summary}
%
%
%
%

\begin{enumerate}
     \item For studying the influence of the photon field on the magnetocentrifugal acceleration of charged particles in rotating magnetospheres we examined dynamics of particles restricted by the photon drag force in the isotropic case. For this purpose three different configurations have been considered: the straight line; the Archimede's spiral and the configuration of the dipolar field line.
     \item
     We found that for straight lines ($\varphi = const$) there is not much difference in dynamics when friction is present. Particle still approaches the light cylinder zone but more slowly as friction coefficient increases. We have shown that the minimum radial velocity when the radial acceleration is initially negative, decreases if the drag force is present.
     \item
     We have found that when the field line has a shape of the Archimede's spiral ($\varphi = a r$) the photon drag force strongly influences dynamics of particles in some cases.
     When $a<0$ and $-\omega /a < 1$ the particle tends to the force free regime: reaction and friction forces cancel each other in infinity. Regardless of the value of friction coefficient the effective angular velocity tends to zero and therefore, the velocity approaches $-\omega /a$.
     When $a>0$ and $\beta > 0$ there is the stable equilibrium point inside the light cylinder. Radial distance of this point decreases as $\beta$ increases.
     \item
     If the field line has a shape of the dipolar magnetic field, $r=L \sin^2 \varphi$, there is always stable equilibrium point. As friction coefficient increases radial distance of equilibrium point decreases. As friction approaches zero this distance tends to furthest point of field line if it is fully in light cylinder and if it isn't than the distance approaches light cylinder radius.

\end{enumerate}

This study was a first attempt of this kind, therefore, in the present paper we only generally considered the problem without particular applications to astrophysical objects. The whole universe is full of the photon field, starting from the cosmic background radiation up to high density photon fields in such extreme environments as the magnetospheres of AGNs and pulsars. In the environments of AGN the magnetic induction is assumed to be of the order of $10^{1-4}$G, which might lead to the frozen-in condition of the plasma particles (Ref. \refcite{AGN}). On the other hand, this medium is rotating and the present study might be easily applicable. In case of pulsars the magnetic field is even more higher, and ranges from $\sim 10^{12}$G to $10^{13}$G (Ref. \refcite{GJ}), which combined with very short periods of rotation ($\sim 0.001$sec-$1$sec) might lead to very interesting results. Therefore, as a next step we intend to apply the methods developed in this work to the mentioned astrophysical objects. A particular interest deserve the stable equilibrium locations, which might be interesting in the context of luminous bunches observed in certain extragalactic jets. Since these knot-like structures we observe as luminous relatively small areas, it is significant to study generation of emission in the mentioned zones. In the framework of our model the particles localised in the aforementioned regions are not entirely static, but, depending on the configuration of the field lines rotate with relativistic velocities with respect to the photon field. Therefore, in the context of the produced emission one should study the back reaction of these particles on the background photon sea, taking into account the inverse Compton processes. On the other hand for generalising the method to jets one has to consider the field lines in 3D.

\section*{Acknowledgments}

The research was supported by the Shota Rustaveli
National Science Foundation grant (DI-2016-14). The research of GB was supported by the Knowledge Foundation at the Free University of Tbilisi.

\end{document}